\documentclass[onecolumn]{IEEEtran}
\usepackage{amsmath}
\usepackage{amsfonts}
\usepackage{latexsym}
\usepackage{amssymb}
\usepackage[noadjust]{cite}
\usepackage[ruled,lined]{algorithm2e}
\usepackage{upref}
\usepackage{theorem}
\usepackage{graphicx}
\usepackage{overpic}
\usepackage{enumitem}
\usepackage{multirow}

\usepackage[usenames]{color}
 \usepackage {tikz}
\usetikzlibrary {positioning}

\usepackage{mathtools}

\DeclarePairedDelimiter\abs{\lvert}{\rvert}
\DeclarePairedDelimiter\norm{\lVert}{\rVert}
\DeclarePairedDelimiter\ceil{\lceil}{\rceil}
\DeclarePairedDelimiter\floor{\lfloor}{\rfloor}
\DeclarePairedDelimiter\parenv{\lparen}{\rparen}

\DeclarePairedDelimiter\set{\{}{\}}
\DeclarePairedDelimiter\mset{\{\!\!\{}{\}\!\!\}}

\theoremstyle{plain}
\newtheorem{theorem}{Theorem}
\newtheorem{lemma}[theorem]{Lemma}

\newtheorem{corollary}[theorem]{Corollary}
\newtheorem{definition}[theorem]{Definition}
\newtheorem{example}[theorem]{Example}
\newtheorem{remark}[theorem]{Remark}
\newtheorem{construction}{Construction}

\newcommand{\cC}{\mathcal{C}}
\newcommand{\cD}{\mathcal{D}}

\newcommand{\cF}{\mathcal{F}}

\newcommand{\cI}{\mathcal{I}}

\newcommand{\cN}{\mathcal{N}}

\newcommand{\cR}{\mathcal{R}}

\renewcommand{\leq}{\leqslant}

\renewcommand{\geq}{\geqslant}

\newcommand{\N}{\mathbb{N}}

\newcommand{\Z}{\mathbb{Z}}

\newcommand{\ba}{\boldsymbol{a}}
\newcommand{\bb}{\boldsymbol{b}}
\newcommand{\bu}{\boldsymbol{u}}
\newcommand{\bv}{\boldsymbol{v}}
\newcommand{\bw}{\boldsymbol{w}}
\newcommand{\bx}{\boldsymbol{x}}
\newcommand{\bX}{\boldsymbol{X}}
\newcommand{\by}{\boldsymbol{y}}
\newcommand{\bz}{\boldsymbol{z}}

\newcommand{\eqdef}{\triangleq}
\newcommand{\sz}{\mathsf{z}}
\newcommand{\emptyseq}{\boldsymbol{\varepsilon}}
\newcommand{\der}{\Longrightarrow}
\DeclareMathOperator{\del}{Del}
\DeclareMathOperator{\rem}{Rem}
\DeclareMathOperator{\rll}{RLL}

\begin{document}


\title{On the Asymptotic Rate of Optimal Codes that Correct Tandem Duplications for Nanopore Sequencing}

\author{
Wenjun Yu, Zuo Ye and Moshe~Schwartz,~\IEEEmembership{Fellow,~IEEE}%
\thanks{Wenjun Yu and Zuo Ye are with the School
   of Electrical and Computer Engineering, Ben-Gurion University of the Negev,
   Beer Sheva 8410501, Israel
   (e-mail: wenjun@post.bgu.ac.il, zuoy@bgu.ac.il).}%
\thanks{Moshe Schwartz is with the Department of Electrical and Computer Engineering, McMaster University, Hamilton, ON, L8S 4K1, Canada, and on a leave of absence from the School
   of Electrical and Computer Engineering, Ben-Gurion University of the Negev,
   Beer Sheva 8410501, Israel
   (e-mail: schwartz.moshe@mcmaster.ca).}
}

\maketitle

\begin{abstract}
We study codes that can correct backtracking errors during nanopore sequencing. In this channel, a sequence of length $n$ over an alphabet of size $q$ is being read by a sliding window of length $\ell$, where from each window we obtain only its composition. Backtracking errors cause some windows to repeat, hence manifesting as tandem-duplication errors of length $k$ in the $\ell$-read vector of window compositions. While existing constructions for duplication-correcting codes can be straightforwardly adapted to this model, even resulting in optimal codes, their asymptotic rate is hard to find. In the regime of unbounded number of duplication errors, we either give the exact asymptotic rate of optimal codes, or bounds on it, depending on the values of $k$, $\ell$ and $q$. In the regime of a constant number of duplication errors, $t$, we find the redundancy of optimal codes to be $t\log_q n+O(1)$ when $\ell|k$, and only upper bounded by this quantity otherwise.
\end{abstract}

\begin{IEEEkeywords}
Nanopore sequencing, tandem duplication, error-correcting codes
\end{IEEEkeywords}
	
\section{Introduction}
\IEEEPARstart{W}{ith} the exponentially increasing amount of data generated and the need for long-term archival data storage, DNA storage is emerging as a contender technology, capable of breaking existing bottlenecks of  conventional electronic storage systems by offering high data density, longevity and ease of copying information~\cite{church2012next}. Among the many challenges facing us in the process of enabling DNA storage, we focus on the sequencing technologies used to read DNA.

Several technologies have been developed to improve the reading performance of DNA. In particular, nanopore sequencing has attracted the attention of researchers and practitioners alike due to better portability, ability to read longer strands, and low cost. The main nanopore sequencing process involves pushing a DNA fragment through a microscopic pore in a lipid membrane. While moving through the pore, a window of length $\ell$ nucleotides (bases) is observed by the machine. As the DNA molecule continues moving through the pore, ideally, we should obtain a sliding-window reading of the entire DNA molecule.

Unfortunately, certain physical aspects of the nanopore sequencer lead to various distortions in the final readout such as random dwell times, fading, inter-symbol interference (ISI), and random noise. Additionally, the passage of the DNA fragment through the pore is often irregular and may involve backtracking or skipping, creating duplicated parts in the sliding-window reading, or missing windows. Finally, it is not always easy to get the window reading, forcing us to resort to weaker forms of reading, for example, only obtaining the (unordered) composition of the bases within each window.

Prior work in this area focused on developing various mathematical models for the sequencer and designing codes to efficiently correct errors in the final readouts. For example, \cite{mao2018models} studied the model that incorporates ISI, deletions and measurement noise. In~\cite{banerjee2024error}, the authors considered codes that correct substitution errors, and provided the optimal redundancy for single substitution with window of length $\ell\geq 3$. The work in~\cite{sun2024bounds} generalized previous works to propose the optimal redundancy for two substitution errors. Additionally, \cite{banerjee2024correcting} studied the single deletion nanopore channel and proposed a code with optimal redundancy up to an additive constant.
  
In this paper, we adopt the nanopore-channel model studied in~\cite{banerjee2024error,sun2024bounds,banerjee2024correcting}: the DNA sequence is read by sliding windows of length $\ell$, where we only obtain the composition of each scanned window. However, unlike~\cite{banerjee2024error,sun2024bounds} which studied substitution errors, and~\cite{banerjee2024correcting} which studied skipping errors (causing deletions), we focus on backtracking errors. In more detail, each DNA sequence of length $n$ is first transformed into an $n+\ell-1$-length vector called an $\ell$-read vector through the ISI channel, where the $i$-th coordinate of the $\ell$-read vector is obtained by collecting the consecutive symbols between the $(i-\ell+1)$-st and $i$-th entries of the original DNA sequence and outputting their (unordered) multiset composition. This process can be regarded as a detecting window of length $\ell$ passing through the whole sequence by shifting one position at each step. However, as the DNA molecule moves through the pore, a backtracking error causes it to move back $k$ bases. Thus, a few of the sliding windows are repeated, in essence, causing a tandem duplication of length $k$ to occur in the $\ell$-read vector.

Duplication errors were introduced in~\cite{FarSchBru16}, in the context of DNA storage in living organisms, and the first duplication-correcting codes were developed in~\cite{jain2017duplication}. These allow for arbitrary alphabets, parametric duplication length, and various duplication types. Following this, codes correcting any number of tandem duplications were studied in~\cite{jain2017duplication,ZerEsmGul19,yu2024duplication}. Codes that correct only a fixed number of tandem duplications (sometimes just one) were studied in~\cite{kovavcevic2019zero,lenz2019duplication,GosPolVor23}. Additionally, codes that correct mixtures of duplications, substitutions, as well as insertions and deletions, have been the focus of~\cite{tang2020single,tang2021error,TanWanLouGabFar23}.

In this paper, we are interested in codes over an alphabet of size $q$ and length $n$, that are capable of correcting tandem duplications of length $k$ in their $\ell$-read vectors. We are interested both in the regime of an unbounded number of duplications, or a constant number of duplications. Since codes capable of correcting tandem duplications exist in both regimes, a simple and effective strategy for constructing the desired codes is the following: Consider the set of all $q^n$ possible sequences of length $n$ over the alphabet. After applying the sliding window, these result in $q^n$ $\ell$-read vectors that reside in a much larger space of size $\sim q^{\ell n}$. Note that most of the sequences in that larger space are \emph{not} $\ell$-read vectors, since they do not represent a sliding-window reading of a sequence over the original alphabet. However, in that large space we can construct a known duplication-correcting code, and from its set of codewords, keep only those that are actually $\ell$-read vectors. Thus, we can harness the power of known optimal constructions, but we are left with a difficult problem: what are the rates of the resulting codes?

The main contribution of this paper is finding the exact asymptotic rate, or bounds on it, for optimal codes over an alphabet of size $q$, that are capable of correcting tandem duplications of length $k$ in their $\ell$-read vectors, both in the regime of unbounded number of errors, and in the regime of a constant number of errors. For the case of unbounded errors, we study the optimal code that parallels the construction given in~\cite{jain2017duplication}. Depending on the values of $k$ and $\ell$, sometimes depending on whether $\ell|k$, we can compute the exact asymptotic rate of the optimal code, and in others, we provide bounds. These results are summarized in Table~\ref{tab:unbounded}. For the case of a constant number of errors, $t$, we study the code that parallels the construction given in~\cite{kovavcevic2018asymptotically}. We show that the redundancy of the optimal code is $t\log_q n + O(1)$ when $\ell|k$, and only upper bounded by this quantity otherwise. All results generalize the results of~\cite{jain2017duplication,kovavcevic2018asymptotically}, which are the special case of $\ell=1$, namely, when sequences are read letter by letter.

The paper is organized as follows. We begin in Section~\ref{sec:prelim} by introducing the relevant notation and definitions. Section~\ref{sec:unbounded} focuses on the regime of unbounded duplication errors, whereas Section~\ref{sec:bounded} studies a constant number of errors. We conclude in Section~\ref{sec:conc} with a short discussion of the results and open problems.

\section{Preliminaries}
\label{sec:prelim}

For any $i,j\in\Z$, $i\leq j$, we define $[i,j]\eqdef \set{i,i+1,\dots,j}$. For $n\in\N$ we then define $[n]\eqdef [1,n]$. Consider some finite alphabet $\Sigma$. A sequence (or, a vector) of length $n$ over $\Sigma$ is an $n$-tuple $\bx=(x_1,x_2,\dots,x_n)$, where $x_i\in\Sigma$ for all $i$. We shall generally use bold lower-case letters for sequences, and the same letter in regular font and an index, to denote a letter of the sequence. Indexing of sequence positions will start from $1$. The length of $\bx$ is denoted by $\abs{\bx}\eqdef n$, and whenever it makes sense, the norm of $\bx$ is defined as $\norm{\bx}_1\eqdef \sum_{i=1}^n \abs{x_i}$. The set of all sequences of length $n$ over $\Sigma$ is denoted by $\Sigma^n$. The unique empty sequence (the sequence of length $0$) is denoted by $\emptyseq$, and we define $\Sigma^0\eqdef\set{\emptyseq}$. We then also define $\Sigma^{\leq n}\eqdef \bigcup_{i=0}^{n}\Sigma^i$, $\Sigma^*\eqdef\bigcup_{i\geq 0}\Sigma^i$, and $\Sigma^+\eqdef\bigcup_{i\geq 1}\Sigma^i$. Given two sequences, $\bx,\by\in\Sigma^*$, we use either $\bx\by$ or $(\bx,\by)$ to denote their concatenation. For any $\ell\in\N$, we also define $\bx^\ell$ to be the concatenation of $\ell$ copies of $\bx$. Naturally, we define $\bx^0\eqdef \emptyseq$.

Consider a sequence $\bx=(x_1,x_2,\dots,x_n)\in\Sigma^n$. We say $\bv\in\Sigma^k$, $k\geq 1$, is a \emph{$k$-factor} of $\bx$, if there exist $\bu,\bw\in\Sigma^*$ such that $\bx=\bu\bv\bw$. If the length of $\bv$ is unknown or unimportant, we just say that $\bv$ is a factor of $\bx$. Given $i,j\in\Z$, $i\leq j$, the window from position $i$ to $j$ is defined as
\[\bx[i,j]\eqdef(x_i, x_{i+1},\dots,x_{j}).\]
To allow for windows that extend outside of $[n]$, we define $x_i\eqdef \emptyseq$ for all $i\notin[n]$. Thus, the factors of $\bx$ are of the form $\bx[i,j]$ for all $1\leq i\leq j\leq n$. We also define the removal of the $[i,j]$ window from $\bx$ by
\[
M_{[i,j]}(\bx) \eqdef (x_1, x_2,\dots,x_{i-1},x_{j+1},x_{j+2},\dots,x_n).
\]
We shall need to count the number of times a specific letter appears in a sequence. Let $a\in\Sigma$ be some letter. Then $\bx|_a$ counts the number of times $a$ appears in $\bx$, namely,
\[
\bx|_a\eqdef \abs*{\set*{i\in[n] ~:~ x_i=a}}.
\]

\begin{example}
Consider the binary sequence $\bx=(0,1,1,0,0,0,1)$. Then $\bx[2,4]=(1,1,0)$, and $M_{[2,4]}(\bx)=(0,0,0,1)$. Also notice that $\bx[-3,2]=(0,1)$ and $\bx[5,12]=(0,0,1)$. Of course, $\bx|_0=4$ since there are four occurrences of $0$ in $\bx$.
\end{example}

Let $\bx\in\Sigma^n$ be a sequence of length $n$, and assume $1\leq m\leq n-1$. We say $\bx$ is $m$-periodic, if $x_{i+m}=x_i$ for all $i\in[1,n-m]$. We note that by saying $\bx$ is $m$-periodic we do not contend that $m$ is the minimal positive integer with this property, but rather merely that it is a period of $\bx$. Thus, we have
\[
\bx = \parenv*{\bx[1,m]^{\ceil{n/m}}}[1,n].
\]
If no such $m$ exists we say that $\bx$ is aperiodic.

In what follows, we shall be examining sequences over various alphabets, which will all be derived from an original sequence. We shall assume, w.l.o.g., that the original sequence is over the alphabet $\Z_q\eqdef\set{0,1,\dots,q-1}$, for some integer $q\geq 2$. Assume, therefore, that $\bx=(x_1,x_2,\dots,x_n)\in\Z_q^n$ is such a sequence. The \emph{composition} of $\bx$ is represented by the polynomial in the variables $\sz_0,\sz_1,\dots,\sz_{q-1}$,
\[
c(\bx) \eqdef \bx|_0\cdot\sz_0 + \bx|_1\cdot\sz_1 +  \dots+ \bx|_{q-1}\cdot\sz_{q-1}.
\]
We note that $c(\emptyseq)=0$. Additionally, if $\bx$ and $\by$ are sequences, then $c(\bx\by)=c(\bx)+c(\by)$.

In the nanopore-sequencing channel setting, a DNA sequence is read by a sliding window of length $\ell$, however, from each such window only its composition is obtained. To that end, we define the following alphabet
\begin{equation}
\label{def:psi}
\Psi_{\ell,q} \eqdef \set*{c(\by) ~:~ \by\in \Z_q^{\leq \ell}}.
\end{equation}
In other words, $\Psi_{\ell,q}$ contains all possible compositions of sequences of length at most $\ell$ over $\Z_q$. Evidently, $\abs{\Psi_{\ell,q}}=\binom{\ell+q}{\ell}$. We then define the composition of the window of length $\ell$ that ends at position $i$ by
\[
R_{i,\ell}(\bx)\eqdef c(\bx[i-\ell+1,i]),\]
for all $i\in[n-\ell+1]$. We also observe that whenever $i\notin[n+\ell-1]$ we have $\bx[i-\ell+1,i]=\emptyseq$, and so $R_{i,\ell}(\bx)=0$. We can now define the \emph{$\ell$-read sequence} of $\bx$ as
\begin{align*}
\cR_\ell(\bx) &\eqdef \parenv*{c(\bx[-\ell+2,1]),c(\bx[-\ell+3,2]),\dots,c(\bx[n,n+\ell-1])}\\
& = (R_{1,\ell}(\bx), R_{2,\ell}(\bx), \dots, R_{n+\ell-1,\ell}(\bx)) \in \Psi_{\ell,q}^{n+\ell-1}.
\end{align*}

\begin{example}
\label{ex:1}
Assume that $q=4$, namely, we are working over $\Z_4$. Let
\[\bx=(1,2,0,1,3,1,2,2,0,0)\in\Z_4^{10}.\]
Take window size $\ell=2$, so then
\[
\cR_2(\bx) = (\sz_1,\sz_1+\sz_2,\sz_0+\sz_2,\sz_0+\sz_1,\sz_1+\sz_3,\sz_1+\sz_3,\sz_1+\sz_2,2\sz_2,\sz_0+\sz_2,2\sz_0,\sz_0)\in\Psi_{2,4}^{11}.
\]
\end{example}

In the nanopore-channel setting, a sequence $\bx$ is transmitted, while $\cR_\ell(\bx)$ is received, assuming a perfect noiseless reading. However, as described in~\cite{laszlo2014decoding}, backtracking errors may occur. When such an error occurs, the nanopore reading mechanism shifts back the molecule in the pore by $k$ positions, therefore repeating the last $k$ entries read, before continuing on the reading of the remainder of the molecule. These backtracking errors are easily described using the tandem-duplication error mechanism already used in~\cite{jain2017duplication} in a different context.

Generally speaking, assume $\by\in\Sigma^*$ is some sequence over the finite alphabet $\Sigma$. If a tandem-duplication error of length $k$ occurs at position $i$, we denote the resulting erroneous sequence by
\[
T_{i,k}(\by) = \bu \bv^2 \bw,
\]
where $\bu,\bv,\bw\in\Sigma^*$, $\by=\bu\bv\bw$, $\abs{\bu}=i$, and $\abs{\bv}=k$. Thus, this error repeats a $k$-factor of $\by$ located right after an $i$-prefix of $\by$. We say $t$ such errors occurred, resulting in the sequence $\bz$, if there exist $i_1,i_2,\dots,i_t$ such that
\[
\bz=T_{i_t,k}\parenv*{T_{i_{t-1},k}\parenv*{\dots T_{i_1,k}(\by)}}.
\]
We denote this relation between $\by$ and $\bz$ by
\[ \by\der^t_k \bz,\]
and say that $\bz$ is a $t$-descendant of $\by$. If the number of errors is finite but unknown or unimportant, we write $\by\der^*\bz$ and say that $\bz$ is a descendant of $\by$. For convenience, $\by$ is a $0$-descendant of itself. The set of all $t$-descendants of $\by$ is denoted by
\[
D^t_k(\by) \eqdef \set*{ \bz\in\Sigma^* ~:~ \by\der^t_k\bz},
\]
and similarly,
\[
D^*_k(\by) \eqdef \set*{ \bz\in\Sigma^* ~:~ \by\der^*_k\bz} = \bigcup_{j=0}^{\infty} D^j_k(\by).
\]
As is common in coding theory, we define the ball of radius $t$ around $\by$ by
\[
B_{k,t}(\by)\eqdef \bigcup_{j=0}^t D^j_k(\by),
\]
and for an unbounded radius, we define
\[
B_{k,\infty}(\by)\eqdef D^*_k(\by).
\]

We now introduce the channel we shall focus on in this paper. In this channel model, a sequence $\bx\in\Z_q^n$ is transmitted by being stored as a DNA sequence. The reading is performed using nanopore sequencing, which ideally, should result in the $\ell$-read vector $\cR_\ell(\bx)$. However, since backtracking errors may occur, tandem duplications occur in $\cR_\ell(\bx)$. We assume these errors are of length $k$. Assuming no more than $t$ such errors occur, the receiver obtains $\bz\in B_{k,t}(\cR_\ell(\bx))$. If the number of backtracking errors is unbounded, $\bz\in B_{k,\infty}(\cR_\ell(\bx))$. In summary,
\begin{equation*}
\bx\in \Z_q^n \xrightarrow{\text{$\ell$-read}} \cR_\ell(\bx) \xrightarrow{\text{tandem duplication}} \bz\in B_{k,\infty}({\cR_\ell (\bx)}).
\end{equation*}

\begin{example}
\label{ex:2}
We continue the setting of Example~\ref{ex:1}. Assume a single backtracking error occurred with length $k=3$, resulting in a tandem duplication of length $3$ in $\cR_2(\bx)$, following a prefix of length $5$,
\[
T_{5,3}(\cR_2(\bx))=(\sz_1,\sz_1+\sz_2,\sz_0+\sz_2,\sz_0+\sz_1,\sz_1+\sz_3,\sz_1+\sz_3,\sz_1+\sz_2,2\sz_2,\underline{\sz_1+\sz_3,\sz_1+\sz_2,2\sz_2},\sz_0+\sz_2,2\sz_0,\sz_0)\in B_{3,1}(\cR_2(\bx)),
\]
where the underline shows the duplicated $3$-factor.
\end{example}

We emphasize that after a tandem duplication, the resulting vector of monomials is not necessarily an $\ell$-read vector of some sequence. Example~\ref{ex:2} illustrates this point since there is no sequence $\bx'$ such that $\cR_2(\bx')=T_{5,3}(\cR_2(\bx))$. In this particular case, this can be seen by the two adjacent entries in $T_{5,3}(\cR_2(\bx))$ containing $2\sz_2,\sz_1+\sz_3$ which can never be adjacent in a $2$-read vector.
  
We shall design error-correcting codes for this channel, which we now define.

\begin{definition}
Let $\ell\in\N$ be the window read size, $k\in\N$ the tandem-duplication length, $t\in\N$ the maximum number of tandem-duplication errors (or $t=\infty$ if this number is unbounded), $n\in\N$ the code length, and $q\in\N$ the alphabet size. A code of length $n$ is a subset $\cC\subseteq\Z_q^n$. We say it can correct $t$ tandem-duplication errors of length $k$ in the $\ell$-read nanopore channel if for any $\bx,\by\in\cC$, $\bx\neq\by$, we have
\[
B_{k,t}(\cR_\ell (\bx)) \cap B_{k,t}(\cR_\ell (\by)) = \emptyset.
\]
The \emph{rate} of the code $\cC$ is defined as 
\[R(\cC)\eqdef \frac{1}{n}\log_q\abs{\cC},\]
and its \emph{redundancy} is defined as $n(1-R(\cC))$.
\end{definition}

\section{Correcting an Unbounded Number of Errors}
\label{sec:unbounded}

In this section we describe an optimal construction for codes that are capable of correcting an unbounded number of tandem duplications in the nanopore channel. The construction parallels the construction of~\cite{jain2017duplication}, in the sense discussed in the introduction. While the construction is straightforward, and even though we can show it is optimal, it is difficult to find its asymptotic rate. The main challenge we face is the fact that our domain is complicated -- the set of $\ell$-read vectors. We manage to find the exact rate for some values of $\ell$ and $k$ (the tandem-duplication length), and bound it for the other cases.

The main tool used in the construction is the $k$-step derivative. It was first employed by~\cite{jain2017duplication} to construct codes correcting any number of tandem duplications, and was later used in~\cite{kovavcevic2018asymptotically,kovavcevic2019zero,lenz2019duplication,yehezkeally2019reconstruction,tang2020single,tang2021error,yu2024duplication}. 

\begin{definition}
\label{def:der}
Assuming $G$ is some (additive) Abelian group, let $\by=(y_1,\dots,y_m)\in G^m$ be a sequence of group elements of length $m$. For $k\geq 1$, the \emph{$k$-step derivative of $\by$} is defined by
\[
\Delta_k(\by) \eqdef (y_1-y_{1-k},y_2-y_{2-k},\dots,y_m-y_{m-k})\in G^m,
\]
where for $i\leq 0$ we define $y_i=0$.
\end{definition}

Building on the definition of $\Psi_{\ell,q}$ from~\eqref{def:psi}, we now define
\[
\Xi_{\ell,q} \eqdef \set*{c(\bx)-c(\by) ~:~ \bx,\by\in \Z_q^{\leq \ell}},
\]
namely, $\Xi_{\ell,q}$ contains the differences of any two polynomials from $\Psi_{\ell,q}$. We shall mostly be interested in the $k$-step derivative of the $\ell$-read vector of a sequence\footnote{Here, the group $G$ from the Definition~\ref{def:der} contains all the linear polynomials in the variables $\sz_0,\dots,\sz_{q-1}$.}. Assume $\bx\in\Z_q^n$ be a sequence, and let $\cR_\ell (\bx)= (R_{1,\ell}(\bx), R_{2,\ell}(\bx), \dots, R_{n+\ell-1,\ell}(\bx))\in\Psi_{\ell,q}^{n+\ell-1}$ be its $\ell$-read vector. Then the $k$-step derivative of $\cR_\ell (\bx)$ is simply
\[
\Delta_k(\cR_\ell(\bx)) \eqdef \parenv*{R_{1,\ell}(\bx)-R_{1-k,\ell}(\bx)  , R_{2,\ell}(\bx)-R_{2-k,\ell}(\bx) , \dots, R_{n+\ell- 1,\ell}(\bx)-R_{n+\ell-k -1,\ell(\bx)}}\in\Xi_{\ell,q}^{n+\ell-1}.
\]

\begin{example}
\label{ex:3}
Let us continue Example~\ref{ex:1} and Example~\ref{ex:2}. Taking the $3$-step derivative of $\cR_2(\bx)$ and $T_{5,3}(\cR_2(\bx))$ we obtain
\begin{align*}
\Delta_3(\cR_2(\bx)) &= (\sz_1,\sz_1+\sz_2,\sz_0+\sz_2,\sz_0,\sz_3-\sz_2,\sz_1+\sz_3-\sz_0-\sz_2,\sz_2-\sz_0,\\
& \qquad 2\sz_2-\sz_1-\sz_3,\sz_0+\sz_2-\sz_1-\sz_3,2\sz_0-\sz_1-\sz_2,\sz_0-2\sz_2),\\
\Delta_3(T_{5,3}(\cR_2(\bx)))&=
(\sz_1,\sz_1+\sz_2,\sz_0+\sz_2,\sz_0,\sz_3-\sz_2,\sz_1+\sz_3-\sz_0-\sz_2,\sz_2-\sz_0,\\
& \qquad 2\sz_2-\sz_1-\sz_3,0,0,0,\sz_0+\sz_2-\sz_1-\sz_3,2\sz_0-\sz_1-\sz_2,\sz_0-2\sz_2).
\end{align*}
In the derivative, the location of the tandem duplication contains a sequence of $0$'s.
\end{example}

We crucially observe that a tandem duplication of length $k$ is equivalent to an injection of the factor $0^k$ into the derivative, at the same location (see Example~\ref{ex:3} for an illustration of this point). This was already noted in~\cite{jain2017duplication}.

Consider a sequence $\by\in\Xi_{\ell,q}^*$. We can always write it in the form
\begin{equation}
\label{eq:zdecom}
\by = \parenv*{0^{m_0},y_1,0^{m_1},y_2,0^{m_2},\dots,y_s,0^{m_s}},
\end{equation}
where for all $i$, $y_i\in \Xi_{\ell,q}\setminus\set{0}$, and $m_i\geq 0$. We then define
\begin{align}
\mu_k(\by) &\eqdef \parenv*{0^{m_0 \bmod k},y_1,0^{m_1\bmod k},y_2,0^{m_2 \bmod k},\dots,y_s,0^{m_s \bmod k}},\nonumber\\
\sigma_k(\by) & \eqdef \parenv*{\floor*{\frac{m_0}{k}},\floor*{\frac{m_1}{k}},\dots,\floor*{\frac{m_s}{k}}}. \label{eq:sigma}
\end{align}
Obviously, $\by$ is easily obtained from $\mu_k(\by)$ and $\sigma_k(\by)$.

\begin{definition}
\label{def:nucleus}
Let $\bx\in\Z_q^n$ be a sequence, and $\ell,k$ positive integers. We define the \emph{nucleus} of $\bx$ (w.r.t.~the parameters $k$ and $\ell$) to be
\[
N_{k,\ell}(\bx)\eqdef \mu_k(\Delta_k(\cR_\ell(\bx))).
\]
\end{definition}

\begin{lemma}
\label{lem:mu}
Let $k$, $\ell$, and $q$ be positive integers. Let $\bz\in\Psi_{\ell,q}^*$, and assume $\bz\der^t_k \bz'$. Then 
\[\mu_k(\Delta_k(\bz'))=\mu_k(\Delta_k(\bz)).\]
Additionally, $\sigma_k(\Delta_k(\bz'))-\sigma_k(\Delta_k(\bz))$ contains only non-negative entries, and
\[
\norm*{\sigma_k(\Delta_k(\bz'))-\sigma_k(\Delta_k(\bz))}_1=t.
\]
\end{lemma}
\begin{IEEEproof}
Consider first a single tandem duplication, namely, $t=1$ and therefore $\bz\der_k \bz'$. By our previous observation, $\Delta_k(\bz')$ differs from $\Delta_k(\bz)$ by an insertion of the factor $0^k$. By definition, $\mu_k(\Delta_k(\bz'))=\mu_k(\Delta_k(\bz))$. Additionally, $\sigma_k(\Delta_k(\bz'))$ differs from $\sigma_k(\Delta_k(\bz))$ by a single entry that has increased by $1$. The claim for general $t$ follows by simple induction.
\end{IEEEproof}

With the last definition in place, we turn to show that for an unbounded number of tandem-duplication errors, the error balls around two sequences intersect if and only if they have the same nucleus.

\begin{lemma}
\label{lem:infintersect}
Let $\bx,\bx'\in\Z_q^*$ be two sequences, and let $k$ and $\ell$ be two positive integers. Then
\[B_{k,\infty}(\cR_\ell (\bx)) \cap B_{k,\infty}(\cR_\ell (\bx')) \neq \emptyset\]
if and only if $N_{k,\ell}(\bx) =  N_{k,\ell}(\bx')$.
\end{lemma}

\begin{IEEEproof}
The proof is essentially the same as~\cite[Lemma 14]{jain2017duplication}. For the first direction, assume that $\bz \in B_{k,\infty}(\cR_\ell (\bx)) \cap B_{k,\infty}(\cR_\ell (\bx'))$. Thus,
\[
\cR_\ell(\bx) \der_k^* \bz \qquad\text{and}\qquad \cR_\ell(\bx') \der_k^* \bz.
\]
By Lemma~\ref{lem:mu},
\[
\mu_k(\Delta_k(\cR_\ell(\bx)))=\mu_k(\Delta_k(\cR_\ell(\bx'))).
\]
Therefore, by definition
\[ N_{k,\ell}(\bx) = N_{k,\ell}(\bx').\]

In the second direction, assume that as in~\eqref{eq:zdecom} and~\eqref{eq:sigma},
\begin{align*}
N_{k,\ell}(\bx) = N_{k,\ell}(\bx') &= (0^{m_0}, y_1, 0^{m_1}, y_2, \dots, y_s, 0^{m_s}),\\
\sigma_k (\Delta_k(\cR_\ell (\bx))) &= (a_0,a_1,\ldots, a_s),\\
\sigma_k (\Delta_k(\cR_\ell (\bx'))) &= (b_0,b_1,\ldots, b_s).
\end{align*}
Then,
\[\Delta_k^{-1}((0^{m_0 + (a_0 + b_0)k}, y_1, 0^{m_1+ (a_1 + b_1)k}, y_2, \dots, y_s, 0^{m_s + (a_s + b_s)k})) \in B_{k,\infty}(\cR_\ell (\bx)) \cap B_{k,\infty}(\cR_\ell (\bx')).\]
\end{IEEEproof}

The code construction follows immediately from this lemma.

\begin{construction}\label{cons:inf}
Let $k$, $\ell$, $n$ and $q\geq 2$ be positive integers. For any two sequences $\bx,\by\in\Z_q^n$, define the equivalence relation $\bx\sim\by$ if and only if $N_{k,\ell}(\bx)=N_{k,\ell}(\by)$. Let $[\bx]$ denote a fixed representative of the equivalence class containing $\bx$. We construct the code:
\[
\cC_{k,\ell}(n) \eqdef \set*{ [\bx] ~:~ \bx\in\Z_q^n},
\]
namely, the code contains exactly one sequence from each of the equivalence classes of $\sim$.
\end{construction}

\begin{corollary}
The code $\cC_{k,\ell}(n)$ from Construction~\ref{cons:inf} is an optimal code capable of correcting any number of tandem duplications of length $k$ in the $\ell$-read vectors of its codewords.
\end{corollary}

\begin{IEEEproof}
The error-correction capability of the code is derived from Lemma~\ref{lem:infintersect} since its codewords have non-intersecting error balls. The code is optimal since it contains a codeword from each possible equivalence class.
\end{IEEEproof}

To better understand the efficiency of the code construction, our next goal is to find the asymptotic rate $\lim_{n\to\infty}R(\cC_{k,\ell}(n))$. The main obstacle here is the fact that the mapping from sequences to $\ell$-read vectors, $\cR_{\ell}:\Z_q^n\to\Psi_{\ell,q}^{n+\ell-1}$, is not always onto (see Example~\ref{ex:3} and the short discussion following it). Further complications stem from the existence of two parameters $k$ and $\ell$ that affect this asymptotic rate. We divide our discussion according to the values of these two parameters.

We begin with the case $\ell=1$. In this case, we note that by identifying $\sz_a$ with $a$, for all $a\in\Z_q$, we get $\cR_1(\bx)=\bx$ and we sidestep the problem of $\cR_\ell$ not being onto. Thus, Construction~\ref{cons:inf} reduces to~\cite[Construction A]{jain2017duplication}, whose asymptotic rate is known~\cite[eq.~(4) and the Appendix]{jain2017duplication}, giving us
\[
\lim_{n\to\infty}R(\cC_{k,1}(n)) = \log_q \lambda_{k-1,q},
\]
where $\lambda_{k,q}$ is the largest root of
\begin{equation}
    \label{eq:gamma}
f_{k,q}(x)\eqdef x^{k+2}-qx^{k+1}+q-1.
\end{equation}
We comment that $\log_q\lambda_{k-1,q}$ is in fact the capacity of the $q$-ary $(0,k-1)$-RLL constrained system (e.g., see~\cite{LinMar85}).

We divide some of the remaining non-trivial cases as follows. First, we study the case: $k=1$ and $\ell \geq 2$ and obtain its asymptotic rate. We continue with the case of $k$ being a multiple of $\ell$. We carefully study the relationship between factors of the form $0^k$ in $\Delta_k(\cR_\ell(\bx))$ and the structure of $\bx$. Finally, we use the probabilistic method to bound the asymptotic rate for the remaining cases.

As we proceed, we shall find the following definition useful.

\begin{definition}
Let $\ell$, $k$ be positive integers. For a sequence $\bx\in\Z_q^*$, we define its \emph{depth} (w.r.t.~the parameters $k$ and $\ell$) as
\[
d_{k,\ell}(\bx) \eqdef \norm*{\sigma_k(\Delta_k(\cR_\ell(\bx)))}_1.
\]
The set of all sequences of depth $i$ will be denoted by
\[
\cD_{k,\ell}^{(i)} \eqdef \set*{\bx\in\Z_q^* ~:~ d_{k,\ell}(\bx)=i}.
\]
\end{definition}

We partition the code from Construction~\ref{cons:inf}, $\cC_{1,\ell}(n)$, according to the depth of the codewords, and define for all $i\geq 0$,
\[
\cC_{k,\ell}^{(i)}(n) \eqdef \cC_{k,\ell}(n) \cap \cD_{k,\ell}^{(i)}.
\]
Since the $\ell$-read vectors are of length $n+\ell-1$, we cannot have any codeword of depth larger than $n+\ell-1$, so we can certainly write
\begin{equation}
\label{eq:partition}
\cC_{k,\ell}(n) = \bigcup_{i=0}^{\floor{\frac{n+\ell-1}{k}}}\cC_{k,\ell}^{(i)}(n).
\end{equation}

\begin{theorem}
\label{th:k1}
Assume that $\ell\geq 2$, and let $\cC_{1,\ell}(n)$ be the code from Construction~\ref{cons:inf}. Then
\[
\lim_{n \to \infty} R(\cC_{1,\ell}(n))  = \log_q (q-1).
\]
\end{theorem}

\begin{IEEEproof}
Fix a value for $n$. 
Let $\bx = (x_1,\ldots, x_n)\in \cC_{1,\ell}^{(i)}(n)$ for some $i\geq 0$. We explicitly compute 
\begin{equation}
\label{eq:d1rl}
\Delta_1(\cR_\ell(\bx)) = (\sz_{x_1}, \dots, \sz_{x_\ell} ~|~ \sz_{x_{\ell+1}}-\sz_{x_1}, \dots, \sz_{x_{n}}-\sz_{x_{n-\ell}} ~|~ -\sz_{x_{n-\ell+1}}, \dots, -\sz_{x_{n-1}}),
\end{equation}
where the purpose of the vertical lines is notational only, allowing us to refer to the first, second, and third parts of $\Delta_1(\cR_\ell(\bx))$. We note that a $0$ may only occur at the middle part of the derivative, exactly when $x_{i} = x_{i-\ell}$. Additionally, the first two parts equal to the $\ell$-step derivative of $\cR_1(\bx)$,
\begin{equation}
\label{eq:dlr1}
\Delta_\ell(\cR_1(\bx)) = (\sz_{x_1}, \dots, \sz_{x_\ell} ~|~ \sz_{x_{\ell+1}}-\sz_{x_1}, \dots, \sz_{x_{n}}-\sz_{x_{n-\ell}}).
\end{equation}

Let us now define 
\[\cN^{(i)} \eqdef \set*{\mu_1(\Delta_\ell(\cR_1(\bx))) ~:~ \bx \in \cC_{1,\ell}^{(i)}(n) }.\]
By~\eqref{eq:d1rl} and~\eqref{eq:dlr1} we have
\[\mu_1(\Delta_1(\cR_\ell(\bx))) = (\mu_1(\Delta_\ell(\cR_1(\bx))),-\sz_{x_{n-\ell+1}}, \dots, -\sz_{x_{n-1}}).\]
It therefore follows that
\begin{equation}
\label{eq:cisandwich}
\abs*{\cN^{(i)}} \leq \abs*{\cC_{1,\ell}^{(i)}(n)} \leq q^{\ell-1}\abs*{\cN^{(i)}}.
\end{equation}

To find the size of $\cN^{(i)}$, we use the same approach as discussed in the case of $\ell=1$. For each $a\in\Z_q$, we identify $\sz_a$ with $a$. Then, the mapping $\Delta_\ell(\cR_1(\cdot)): \Z_q^n \rightarrow \Z_q^n$ is bijective, and we have 
\[\cN^{(i)} = \set*{\by=(y_1,\dots,y_{n-i}) \in \Z_q^{n-i} ~:~  y_{j}\neq 0 \text{ for } \ell+1 \leq j \leq n-i.}\]
Thus,
\[\abs*{\cN^{(i)}} = q^\ell(q-1)^{n-\ell-i} \leq q^\ell(q-1)^{n-\ell} = \abs*{\cN^{(0)}},\]
and, by~\eqref{eq:cisandwich},
\[\abs*{\cN^{(0)}} \leq \abs*{\cC_{1,\ell}^{(0)}(n)}\leq \abs*{\cC_{1,\ell}(n)} = \sum\limits_{i=0}^{n+\ell-1} \abs*{\cC_{1,\ell}^{(i)}(n)} \leq (n+\ell) q^{\ell-1} \abs*{\cN^{(0)}}.\]
It follows that
\[\lim_{n\to \infty} R(\cC_{1,\ell}(n)) = \lim_{n\to \infty} \frac{\log_q \abs*{\cN^{(0)}}}{n} = \log_q (q-1).\]
\end{IEEEproof}

We continue with the remaining cases, namely, when $k,\ell\geq 2$. In order to bound the size $\cC_{k,\ell}(n)$, we look at its partition into $\cC_{k,\ell}^{(i)}$. In turn, to bound the size of $\cC_{k,\ell}^{(i)}$ we need to better understand the number of sequences $\bx$ with a prescribed number of occurrences of the factor $0^k$ in $\Delta_k(\cR_\ell(\bx))$. Let us first consider the existence of a single $0$ at position $i+\ell+k-1$ of $\Delta_k(\cR_\ell(\bx))$. This position contains $R_{i+\ell+k-1,\ell}(\bx)-R_{i+\ell-1,\ell}(\bx)$. We note that the subtracted value, $R_{i+\ell-1,\ell}(\bx)$, is the composition of $\bx[i,i+\ell-1]$. If the result of the subtraction is to be $0$, and assuming a long enough sequence, we must have $i\geq 1$, or else the subtracted window overlaps only partially with $\bx$. Thus, in order to get a $0$, we must have
\begin{equation}
\label{eq:0k}
R_{i+\ell-1,\ell}(\bx) = R_{i+\ell+k-1,\ell}(\bx).
\end{equation}
Now, to get a factor $0^k$ in $\Delta_k(\cR_\ell(\bx))$, we need to require~\eqref{eq:0k} for $k$ consecutive indices, namely,
\[
R_{i+\ell-1+j,\ell}(\bx) = R_{i+\ell+k-1+j,\ell}(\bx)
\text{ for all } j\in[0,k-1].
\]
The implications of this are detailed in the following lemma.

\begin{lemma}
\label{lem:0k}
Let $k,\ell \geq 2$ and $i\geq 1$ be positive integers. Let $\bx = (x_1,\dots, x_n) \in \Z_q^n$. If $0^k$ occurs starting at position $i+\ell+k-1$ of $\Delta_{k} (\cR_\ell(\bx))$, then one of the following holds:
\begin{enumerate}
\item
$\ell\leq k$, and there exists a subset $\cI\subseteq[0,\ell-1]$ such that for all $r\in[0,\floor{\frac{k+\ell-s-2}{\ell}}]$,
\[
\begin{cases}
x_{i+s+r\ell}=x_{i+k+s+r\ell} & \text{if $s\in\cI,$} \\
x_{i+s}=x_{i+s+r\ell} \text{ and } x_{i+k+s}=x_{i+k+s+r\ell} & \text{if $s\in[0,\ell-1]\setminus\cI$.}
\end{cases}
\]
If additionally $\ell|k$, then only $\cI=[0,\ell-1]$ is possible, and $\bx[i,i+2k+\ell-2]$ is $k$-periodic.
\item
$\ell> k$, and there exists a subset $\cI\subseteq[0,k-2]$ such that,
\[
\begin{cases}
x_{i+s}=x_{i+\ell+s} \text{ and } x_{i+k+s}=x_{i+k+\ell+s} & \text{if $s\in\cI$,}\\
x_{i+s}=x_{i+k+s} \text{ and } x_{i+\ell+s}=x_{i+k+\ell+s} & \text{if $s\in[0,k-2]\setminus\cI$.}
\end{cases}
\]
\end{enumerate}
\end{lemma}

\begin{IEEEproof}
Assume that $0^k$ occurs starting at position $i+\ell+k-1$ of $\Delta_k(\cR_\ell(\bx))$. We then have
\begin{equation}
\label{eq:req}
R_{i+\ell-1+j,\ell}(\bx) = R_{i+k+\ell - 1+j,\ell}(\bx) \text{ for all } j\in[0,k-1].
\end{equation}
These equalities involve the symbols of $\bx[i,i+2k+\ell-2]$. Since
\begin{align*}
R_{i+\ell-1+j,\ell}(\bx) =c(x_{i+j}, \ldots, x_{i+\ell-1+j}) &= c(x_{i+k+j}, \ldots, x_{i+k+\ell-1+j}) = R_{i+k+\ell-1+j,\ell}(\bx),\\
R_{i+\ell+j,\ell}(\bx) =c(x_{i+1+j}, \ldots, x_{i+\ell+j}) &= c(x_{i+k+1+j}, \ldots, x_{i+k+\ell+j}) = R_{i+k+\ell+j,\ell}(\bx),
\end{align*}
for any $j\in[0,k-2]$, we have 
\begin{equation}
\label{eq:pair}
\mset*{x_{i+j}, x_{i+k+\ell+j}} = \mset*{x_{i+\ell+j},x_{i+k+j}},
\end{equation}
where $\mset{\dots}$ denotes a multiset. We divide our proof depending on $\ell$ and $k$.

\textbf{Case 1:} $\ell\leq k$. We define the following set of integers:
\[
\cI \eqdef \set*{s\in [0,\ell-1] ~:~ x_{i+s} = x_{i+k+s}}
\]
By using~\eqref{eq:pair}, we obtain the following:
\begin{enumerate}
\item 
For all $s\in \cI$, $x_{i+s}=x_{i+k+s}$, which implies $x_{i+\ell+s} = x_{i+k+\ell+s}$. 
\item 
For all $s\in [0,\ell-1] \setminus\cI$, $x_{i+s}\neq x_{i+k+s}$, which implies $x_{i+s} =  x_{i+\ell+s}$ and $x_{i+k+s} =  x_{i+k+\ell+s}$. We also must have $x_{i+\ell+s} \neq x_{i+k+\ell+s}$. 
\end{enumerate}
We can iterate this process, and get:
\begin{enumerate}[label=$\alph*$)]
\item \label{(a)}
For all $s\in \cI$, $x_{i+s} = x_{i+k+s}$ and $x_{i+s+r\ell} =  x_{i+k+s+r\ell}$ for all $1 \leq r \leq \floor{\frac{k+\ell-s-2}{\ell}}$.
\item \label{(b)}
For all $s\in [0,\ell-1] \setminus \cI$, $x_{i+s} =  x_{i+s+r\ell}$ and $x_{i+k+s} =  x_{i+k+s+r\ell}$ for all $1 \leq r \leq \floor{\frac{k+\ell-s-2}{\ell}}$.
\end{enumerate}

In addition, we consider the sub-case of $\ell|k$. Assume to the contrary that $\cI \neq [0,\ell-1]$. Recall, from~\eqref{eq:req}, that $R_{i+\ell-1,\ell}(\bx)=R_{i+k+\ell-1,\ell}(\bx)$. This means that $x_{i},\dots,x_{i+\ell-1}$ is a permutation of $x_{i+k},\dots,x_{i+k+\ell-1}$. Thus, if $\cI\neq[0,\ell-1]$, we cannot have $\abs{\cI}=\ell-1$, and we must have $\abs{\cI}\leq \ell-2$. Thus, there exists $s\in[0,\ell-2]$ such that $x_{i+s} \neq x_{i+k+s}$. This, however, contradicts $\ref{(b)}$ since we can choose $r=\frac{k}{\ell}\leq \floor{\frac{k+\ell-s-2}{\ell}}$, and we get $x_{i+s}=x_{i+s+rl}=x_{i+k}$.

Thus, when $\ell|k$ we have $\cI=[0,\ell-1]$. But then, \ref{(a)} implies that $x_{i+j+k}=x_{i+j}$ for all $j\in[0,k+\ell-2]$. Hence, $\bx[i,i+2k+\ell-2]$ is $k$-periodic.

\textbf{Case 2:} $\ell> k$. We once again look at~\eqref{eq:req}. Taking $j=0$,
\[R_{i+\ell-1,\ell}(\bx)=c(x_{i},\dots,x_{i+\ell-1})=c(x_{i+k},\dots,x_{i+k+\ell-1})=R_{i+k+\ell-1,\ell}(\bx).\]
Since $\ell\geq k+1$, we can omit the common part $x_{i+k},\dots,x_{i+\ell-1}$ and obtain,
\[
\mset*{x_{i},\dots,x_{i+k-1}}=\mset*{x_{i+\ell},\dots,x_{i+\ell+k-1}}.
\]
Let us now define
\[\cI \eqdef \set*{s\in [0,k-2] ~:~ x_{i+s} = x_{i+\ell+s}},\]
where we emphasize the fact that we do not consider the case of $s=k-1$. Therefore, by using~\eqref{eq:pair}:
\begin{enumerate}
\item 
For all $s\in \cI$, $x_{i+s} = x_{i+\ell+s}$ and $x_{i+k+s} =  x_{i+k+\ell+s}$.
\item 
For all $s\in [0,k-2] \setminus \cI$, $x_{i+s} =  x_{i+k+s}$ and $x_{i+\ell+s} =  x_{i+k+\ell+s}$.
\end{enumerate}
This completes all the cases of the theorem.
\end{IEEEproof}

Lemma~\ref{lem:0k} shows the importance of sequences with no periodic factor, with proper parameters $k$ and $\ell$. We formalize such sequences in the following definition.

\begin{definition}
Let $k,\ell\geq 2$, and let $\bx\in\Z_q^n$ be a sequence. We say that $\bx$ is \emph{$(k,\ell)$-fine} if it does not contain any $k$-periodic factor of length $2k+\ell-1$. Namely, either the sequence is short $n\leq 2k+\ell-2$, or $n\geq 2k+\ell-1$ and for all $1\leq i\leq n-2k-\ell+2$, $\bx[i,i+2k+\ell-2]$ is not $k$-periodic. We denote the set of all $(k,\ell)$-fine sequences of length $n$ by $\cF_{k,\ell}(n)$.
\end{definition}

Some simple monotonicity is easy to show.

\begin{lemma}
\label{lem:Fn}
Let $k,\ell\geq 2$, and $n\geq 2k+\ell-1$. Then
\[
\abs*{\cF_{k,\ell}(n)}\leq \abs*{\cF_{k,\ell}(n+1)}.
\]
\end{lemma}
\begin{IEEEproof}
We simply show that any sequence in $\cF_{k,\ell}(n)$ can be extended by one symbol to a sequence in $\cF_{k,\ell}(n+1)$. Let $\bx\in\cF_{k,\ell}(n)$, and let $a\in\Z_q$ be a symbol such that $a\neq\bx[n-k+1]$ (which is always possible since $q\geq 2$, namely, the alphabet contains at least two letters). We contend that $\bx a\in\cF_{k,\ell}(n+1)$. Obviously any factor of length $2k+\ell-1$ of $\bx a$ that does not involve the last symbol, is not $k$-periodic since $\bx\in\cF_{k,\ell}(n)$. The only factor in question is therefore $\bx[n-2k-\ell+3,n]a$, but since $a\neq \bx[n-k+1]$, it is certainly not $k$-periodic.
\end{IEEEproof}

Using these tools, we can now bound the rate of the code we constructed.

\begin{theorem}
\label{thm:infbound}
Let $k,\ell\geq 2$, and let $\cC_{k,\ell}(n)$ be the code from Construction~\ref{cons:inf}. Then
\[\lim_{n\to \infty}R(\cC_{k,\ell}^{(0)}(n)) \leq \lim_{n\to \infty}R(\cC_{k,\ell}(n)) \leq \lim_{n\to \infty}R(\cF_{k,\ell}(n)).\] 
Moreover, if $\ell|k$ then, 
\[\lim_{n\to \infty}R(\cC_{k,\ell}(n))  = \lim_{n\to \infty}R(\cC_{k,\ell}^{(0)}(n)) = \lim_{n\to \infty}R(\cF_{k,\ell}(n)).\]
\end{theorem}

\begin{IEEEproof}
The lower bound is trivial since $\cC_{k,\ell}^{(0)}(n) \subseteq \cC_{k,\ell}(n)$. We therefore turn to prove the upper bound. We first contend that $\cC_{k,\ell}^{(0)}\subseteq \cF_{k,\ell}(n)$ when $n\geq 2k+\ell-1$. Assume to the contrary that is not the case, namely, that there exists $\bx\in\cC_{k,\ell}^{(0)}(n)$ such that $\bx\not\in\cF_{k,\ell}(n)$. Thus, $\bx$ contains a factor of length $2k+\ell-1$ that is $k$-periodic, say, $\bx[i,i+2k+\ell-2]$. However, that would mean that $\Delta_k(\cR_\ell(\bx))$ contains $0^k$ starting at position $i$, which contradicts the definition of $\cC_{k,\ell}^{(0)}$. Thus, $\cC_{k,\ell}^{(0)}(n)\subseteq\cF_{k,\ell}(n)$ and then 
\[\abs*{\cC_{k,\ell}^{(0)}(n)}\leq\abs*{\cF_{k,\ell}(n)}.\]

Now, let us consider $\bx \in \cC_{k,\ell}^{(i)}(n)$ for $i\geq 1$. If $\bx$ is $(k,\ell)$-fine, namely, then $\bx \in \cF_{k,\ell}(n)$. Otherwise, $\bx$ has a factor of length $2k+\ell-1$ that is $k$-periodic. Assume the first such factor is  $(x_i,\dots, x_{i+2k+\ell-2})$. We then define $\del_{k,\ell}(\bx)$ as 
\[\del_{k,\ell}(\bx) \eqdef 
M_{[i,i+k-1]}(\bx) = \set{x_1,\ldots,x_{i-1},x_{i+k},\ldots, x_n},\]
namely, $\del_{k,\ell}(\bx)$ deletes the first $k$ symbols of the first factor of length $2k+\ell-1$ that is $k$-periodic. We call this operation a \emph{$k$-periodic deletion at position $i$}. 

Turning our focus to $\Delta_k(\cR_\ell(\bx))$, the $k$-periodic factor of length $2k+\ell-1$ at position $i$ of $\bx$, manifests as a $0^k$ factor at position $i+k+\ell-1$ at $\Delta_k(\cR_\ell(\bx))$. Let us define $\rem_{k,\ell}(\Delta_k(\cR_\ell(\bx)))$ as the operator removing the said $0^k$ factor. Thus,
\begin{align*}
\rem_{k,\ell}(\Delta_k(\cR_\ell(\bx))) &= M_{[i+k+\ell-1,i+2k+\ell-2]}(\Delta_k(\cR_\ell(\bx))) \\
&= (R_{1,\ell}(\bx) - R_{1-k,\ell}(\bx), \dots, R_{i+k+\ell-2,\ell}(\bx)- R_{i+\ell-2,\ell}(\bx), \\
&\qquad R_{i+2k+\ell-1,\ell}(\bx)- R_{i+k+\ell-1,\ell}(\bx),\dots, R_{n+\ell-1,\ell}(\bx)- R_{n+\ell-k-1,\ell}(\bx)).
\end{align*}
We call this the \emph{$0^k$ removal at position $i+k+\ell-1$}.

We contend that
\begin{equation}
\label{eq:commute}
\rem_{k,\ell}(\Delta_k(\cR_\ell(\bx)))=\Delta_k(\cR_\ell(\del_{k,\ell}(\bx))).
\end{equation}
Namely, that deleting the first factor of $\bx$ that has length $2k+\ell-1$ and is $k$-periodic (which is responsible for a $0^k$ factor in the derivative) and then taking the $k$-step derivative of the $\ell$-read vector, is the same as first taking the $k$-step derivative of the $\ell$-read vector and then removing the said $0^k$ factor. The proof of this contention is by brute-force inspection: For convenience, denote
\begin{align*}
\cR_\ell(\bx)&=(R_1,R_2,\dots,R_{n+\ell-1}),\\
\cR_\ell(\del_{k,\ell}(\bx))&=(R'_1,R'_2,\dots,R'_{n+\ell-k-1}).
\end{align*}
Then:
\begin{itemize}
\item 
$R'_j \overset{(a)}{=} R_j$ for $1\leq j\leq i-1$,
\item 
$R'_j = c(x_{j-\ell+1},x_{j-\ell+2},\cdots, x_{i-1},x_{i+k},\ldots, x_{j+k}) \overset{(b)}{=} c(x_{j-\ell+1},x_{j-\ell+2},\cdots, x_{i-1},x_{i},\ldots, x_{j}) = R_j$ for $i\leq j\leq i+\ell-2$,
\item 
$R'_j \overset{(c)}{=} R_{j}$ for $i+\ell -1 \leq j\leq i+k+\ell-2$,
\item 
$R'_j \overset{(d)}{=} R_{j+k}$ for $i+\ell -1 \leq j\leq n+\ell-k-1$,
\end{itemize}
where $(a),(d)$ come from the definition and $(b),(c)$ come from the form of the $k$-periodic factor $(x_i,\ldots,x_{i+2k+\ell-2})$. Hence:
\begin{itemize}
\item
$R'_{j}- R'_{j-k} = R_{j} - R_{j-k}$ for $1 \leq s\leq i+k+\ell-2$,
\item 
$R'_{j} - R'_{j-k} = R_{j+k} - R_{j}$ for $i+k+\ell -1 \leq j \leq n+\ell-k-1$,
\end{itemize}
thus proving our contention.

An important consequence of~\eqref{eq:commute} is that the nucleus of $\bx$ (see Definition~\ref{def:nucleus}) remains unchanged under $\del_{k,\ell}(\cdot)$, namely,
\[
N_{k,\ell}(\del_{k,\ell}(\bx))=N_{k,\ell}(\bx).
\]
We can therefore repeatedly apply $\del_{k,\ell}$ until we reach a $(k,\ell)$-fine sequence. We denote that sequence by $\del_{k,\ell}^*(\bx)$. Obviously,
\begin{equation}
\label{eq:ndel}
N_{k,\ell}(\del^*_{k,\ell}(\bx))=N_{k,\ell}(\bx).
\end{equation}
Additionally,
\begin{equation}
\label{eq:delf}
\del^*_{k,\ell}(\bx)\in \bigcup_{j=0}^{\floor{n/k}} \cF_{k,\ell}(n-jk).
\end{equation}

We observe that
\begin{align*}
\abs*{\cC_{k,\ell}^{(i)}(n)}&\overset{(a)}{=}\abs*{\set*{N_{k,\ell}(\bx) ~:~ \bx\in\cC_{k,\ell}^{(i)}(n)}} \\
&\overset{(b)}{=} \abs*{\set*{N_{k,\ell}(\del_{k,\ell}^*(\bx)) ~:~ \bx\in\cC_{k,\ell}^{(i)}(n)}} \\
& \overset{(c)}{\leq} \abs*{\set*{\del_{k,\ell}^*(\bx) ~:~ \bx\in\cC_{k,\ell}^{(i)}(n)}}\\
&\overset{(d)}{\leq} \sum_{j=0}^{\floor{n/k}}\abs*{\cF_{k,\ell}(n-jk)}\\
&\overset{(e)}{\leq} n \abs{\cF_{k,\ell}(n)},
\end{align*}
where $(a)$ follows from the code definition in Construction~\ref{cons:inf}, $(b)$ follows from~\eqref{eq:ndel}, $(c)$ follows from the fact that several sequences may have the same nucleus, $(d)$ follows from~\eqref{eq:delf}, and $(e)$ follows from Lemma~\ref{lem:Fn} and $k\geq 2$. Thus, by also using~\eqref{eq:partition},
\[
\abs*{\cC_{k,\ell}(n)} = \sum_{i=0}^{\floor{\frac{n+\ell-1}{k}}} \abs*{\cC_{k,\ell}^{(i)}(n)} \leq n^2 \abs{\cF_{k,\ell}(n)}.
\]
Therefore,
\[\lim_{n\to \infty}R(\cC_{k,\ell}(n)) \leq \lim_{n\to \infty}R(\cF_{k,\ell}(n)),\] 
as claimed.

Finally, assume the case of $\ell|k$. By Lemma~\ref{lem:0k}, every factor $0^k$ of $\Delta_k(\cR_\ell(\bx))$ implies the existence of a $k$-periodic factor of $\bx$ of length $2k+\ell-1$. Thus, not only $\cC_{k,\ell}^{(0)}\subseteq\cF_{k,\ell}(n)$ as in the general case, but necessarily $\cC_{k,\ell}^{(0)}=\cF_{k,\ell}(n)$. By sandwiching we get
\[
\lim_{n\to \infty}R(\cC_{k,\ell}(n)) = \lim_{n\to \infty}R(\cC_{k,\ell}^{(0)}(n)) = \lim_{n\to \infty}R(\cF_{k,\ell}(n)),
\]
which completes the proof.
\end{IEEEproof}

We proceed to obtain closed-form bounds on the asymptotic rate of $\cC_{k,\ell}(n)$ from Construction~\ref{cons:inf}. For the upper bound, we note that $\cF_{k,\ell}(n)$ is in fact related to run-length-limited (RLL) sequences, whose asymptotic rate is already known. As for the lower bound, we use the probabilistic method, and in particular, the Lov\'{a}sz Local Lemma in one of its many forms.

\begin{lemma}[{{\cite[Lemma 5.1.1]{alon2016probabilistic}}}]
\label{lem:local}
Let $A_1, \dots, A_m$ be events in an arbitrary probability space. Let $G=(V,E)$ be a directed graph with $V=[m]$ such that for every $i\in [m]$, the event $A_i$ is mutually independent of all the events $\set{A_j: (i,j)\not\in E}$. Suppose that there are real numbers $c_1,\dots,c_m$ such that $0\leq c_i < 1$ and $\Pr(A_i)\leq c_i \prod_{(i,j)\in E}(1-c_j)$ for all $i\in [m]$. Then
\[\Pr\parenv*{\bigwedge_{i\in [m]} \overline{A_i}}\geq \prod_{i\in [m]} (1 - c_i).\]
\end{lemma}

Before stating the theorem, we need another technical lemma.

\begin{lemma}
\label{lem:upperfactor}
Let $k,\ell \geq 2$ and $i\geq 1$ be positive integers. Let $\bx = (x_1,\dots, x_n) \in \Z_q^n$. If $0^k$ occurs starting at position $i+\ell+k-1$ of $\Delta_{k} (\cR_\ell(\bx))$, then the number of possible values for $\bx[i,i+2k+\ell-2]$ is upper bounded by
\[
\leq \begin{cases}
\ell!q^k & \text{if $\ell\leq k$,}\\
k!q^\ell & \text{if $\ell>k$.}
\end{cases}
\]
\end{lemma}
\begin{IEEEproof}
Assume that $0^k$ occurs starting at position $i+\ell+k-1$ of $\Delta_k(\cR_\ell(\bx))$. We then have
\begin{equation}
\label{eq:rs}
R_{i+\ell-1+j,\ell}(\bx) = R_{i+k+\ell - 1+j,\ell}(\bx) \text{ for all } j\in[0,k-1],
\end{equation}
which involves exactly the symbols of $\bx[i,i+2k+\ell-2]$. We first contend that if we know the value of $\bx[i,i+k+\ell-1]$, then there is exactly one way of completing $\bx[i+k+\ell,i+2k+\ell-2]$ such that~\eqref{eq:rs} holds. Indeed, since $R_{i+\ell,\ell}(\bx)=R_{i+\ell+k,\ell}(\bx)$ by~\eqref{eq:rs}, we have
\begin{align*}
\mset*{x_{i+k+\ell}}&=\parenv*{\mset*{x_{i+k}}\cup\mset*{x_{i+k+1}\dots,x_{i+k+\ell}}}\setminus
\mset*{x_{i+k},\dots,x_{i+k+\ell-1}}\\
&= \parenv*{\mset*{x_{i+k}}\cup\mset*{x_{i+1}\dots,x_{i+\ell}}}\setminus
\mset*{x_{i+k},\dots,x_{i+k+\ell-1}},
\end{align*}
which determines the value of $x_{i+k+\ell}$ using the known values of $\bx[i,i+k+\ell-1]$. Iterating this by increasing the indices, we determine the values of $\bx[i+k+\ell,\dots,i+2k+\ell-2]$. We therefore focus on bounding the number of values $\bx[i,i+k+\ell-1]$ can take.

For the first case, assume $\ell\leq k$. There are $q^k$ ways of choosing $\bx[i,i+k-1]$. By~\eqref{eq:rs}, $R_{i+\ell-1,\ell}(\bx)=R_{i+k+\ell-1,\ell}(\bx)$, so
\[
\mset*{x_i,\dots,x_{i+\ell-1}}=\mset*{x_{i+k},\dots,x_{i+k+\ell-1}}.
\]
Since $\ell\leq k$, the two multisets do not overlap. Thus, $\bx[i,i+\ell-1]$ is the same as $\bx[i+k,i+k+\ell-1]$ up to a permutation of the $\ell$ symbols. There are $\ell!$ ways of choosing a permutation, and we obtain the claimed upper bound, $\ell!q^k$.

For the second case, assume $\ell> k$. There are $q^\ell$ ways of choosing $\bx[i,i+\ell-1]$. By~\eqref{eq:rs}, $R_{i+\ell-1,\ell}(\bx)=R_{i+k+\ell-1,\ell}(\bx)$, so
\[
\mset*{x_i,\dots,x_{i+\ell-1}}=\mset*{x_{i+k},\dots,x_{i+k+\ell-1}}.
\]
Because $\ell>k$, the two multisets overlap in $\mset{x_{i+k},\dots,x_{i+\ell-1}}$. Removing this overlap we get
\[
\mset*{x_{i},\dots,x_{i+k-1}}=\mset*{x_{i+\ell},\dots,x_{i+k+\ell-1}}.
\]
Thus, $\bx[i,i+k-1]$ is the same as $\bx[i+\ell,i+k+\ell-1]$ up to a permutation of the $k$ symbols. There are $k!$ ways of choosing a permutation, and we obtain the claimed upper bound, $k!q^\ell$.
\end{IEEEproof}

\begin{theorem}
\label{thm:upbound}
Let $k,\ell\geq 2$, and let $\cC_{k,\ell}(n)$ be the code from Construction~\ref{cons:inf} for alphabet size $q\geq 2$. Then
\[\lim_{n\to \infty}R(\cC_{\ell,k}(n)) \leq \log_q \lambda_{k+\ell-2,q}.\]
If additionally $\ell|k$ then
\[\lim_{n\to \infty}R(\cC_{\ell,k}(n)) = \log_q \lambda_{k+\ell-2,q},\]
where $\lambda_{k+\ell-2,q}$ is defined in~\eqref{eq:gamma}.
\end{theorem}	
 
\begin{IEEEproof} 
Our starting point is the upper bound presented in Theorem~\ref{thm:infbound}. It is shown using the asymptotic rate of $\cF_{k,\ell}(n)$. Here, we simply find the asymptotic rate of $\cF_{k,\ell}(n)$. Using the same approach as before, we take any $\bx\in\Z_q^n$ and look at $\Delta_k(\cR_1(\bx))$. Since we use the $1$-read vector of $\bx$, we identify the monomial $\sz_a$ with the element $a$, for any $a\in\Z_q$. Thus, $\Delta_k(\cR_1(\bx))$ is just a sequence over $\Z_q$ as well. We also note that $\bx$ contains a factor of length $2k+\ell-1$ that is $k$-periodic, if and only if $\Delta_k(\cR_1(\bx))$ contains a factor $0^{k+\ell-1}$. Thus, $\cF_{k,\ell}(n)$ contains exactly those sequences whose $k$-step derivative of their $1$-read vector does not contain the factor $0^{k+\ell-1}$.

The sequences in $\Z_q^n$ that do not contain the factor $0^{k+\ell-1}$ are called $q$-ary $(0,k+\ell-2)$-run-length-limited (RLL) sequences, and are denoted by $\rll_q(0,k+\ell-2)(n)$. Since there is a clear bijection between $\Z_q^n$ and its image under $\Delta_k(\cR_1(\cdot))$ we get
\[
\abs*{\cF_{k,\ell}(n)}=\abs*{\rll_q(0,k+\ell-2)(n)}.
\]
Thus,
\[
\lim_{n\to_\infty}R(\cF_{k,\ell}(n))=\lim_{n\to\infty}R(\rll_q(0,k+\ell-2)(n))=\log_q \lambda_{k+\ell-2,q},
\]
where the last equality is from~\cite[(4) and the Appendix]{jain2017duplication}.

If $\ell|k$ then by Theorem~\ref{thm:infbound} the asymptotic rate of $\cC_{k,\ell}(n)$ equals the asymptotic rate of $\cF_{k,\ell}(n)$, which equals the asymptotic rate of $\rll_q(0,k+\ell-2)(n)$.
\end{IEEEproof}

\begin{remark}
\label{rem:alpha}
We can give more pleasant upper and lower bounds on $\log_q \lambda_{k+\ell-2,q}$ that appears in Theorem~\ref{thm:upbound}. By~\cite[(4) and the Appendix]{jain2017duplication} we can write
\[1 - \frac{\alpha(q-1) \log_q e}{q^{k+\ell}}\leq \log_q \lambda_{k+\ell-2,q} \leq 1 - \frac{(q-1) \log_q e}{q^{k+\ell}},\]
where
\[
\alpha=e^{-W\parenv*{-\frac{q-1}{q^{k+\ell}}(k+\ell)}},
\]
and $W(\cdot)$ is the Lambert $W$-function. As $k+\ell\to\infty$ we have $\alpha\to 1$.
\end{remark}

\begin{theorem}
\label{thm:lowbound}
Let $k,\ell\geq 2$, and let $\cC_{k,\ell}(n)$ be the code from Construction~\ref{cons:inf}. Then
\[\lim_{n\to \infty}R(\cC_{\ell,k}(n)) \geq 1 - \log_q \parenv*{\frac{1}{1-c}}.\]
If $\ell\leq k$, then
\[c = \frac{1 - \sqrt{1 - \frac{\ell!(16k+8\ell-16)}{q^{k+\ell-1}}}}{8k+4\ell-8},\]
and we require
\[q\geq \ceil*{(\ell!(16k+8\ell-16))^{\frac{1}{k+\ell-1}}}.\] 
If $\ell>k$,
\[c = \frac{1 - \sqrt{1 - \frac{k!(16k+8\ell-16)}{q^{2k-1}}}}{8k+4\ell-8},\]
and we require
\[q\geq \ceil*{(k!(16k+8\ell-16))^{\frac{1}{2k-1}}}.\] 
\end{theorem}	
 
\begin{IEEEproof} 
Our starting point, once again, is the lower bound presented in Theorem~\ref{thm:infbound}, which is shown using the asymptotic rate of $\cC_{k,\ell}^{(0)}(n)$. We lower bound this asymptotic rate by using the Lov\'asz Local Lemma.

Let $\bX=(X_1,\dots,X_n)\in\Z_q^n$ be a random sequence, where for each $i\in[1,n]$, $X_i$ is i.i.d. uniformly at random chosen from $\Z_q$. Thus, $\bX$ is uniformly at random chosen from $\Z_q^n$. Let $A_i$ be the event that $\Delta_k(\cR_\ell(\bX))$ has $0^k$ at position $i+\ell+k-1$, where $1\leq  i\leq n-2k-\ell+2.$ Let $G=(V,E)$ be a directed graph with $V=[n-2k-\ell+2]$ and $(i,j)\in E$ if and only if $\abs{i-j}\leq 2k+\ell-2$. By the observation in the proof of Lemma~\ref{lem:upperfactor}, $A_i$ depends on $\bX[i,i+2k+\ell-2]$ only, and so the event $A_i$ is mutually independent of all the events $\set{A_j ~:~ (i,j)\not\in E}$. 

By Lemma~\ref{lem:upperfactor},
\begin{equation}
\label{eq:prup}
\Pr(A_i) \leq 
\begin{cases}
\frac{\ell ! q^{k}}{q^{2k+\ell-1}} = \frac{\ell!}{q^{k+\ell-1}} & \text{if $\ell\leq k$},\\
\frac{k ! q^{\ell}}{q^{2k+\ell-1}} = \frac{k!}{q^{2k-1}} & \text{if $\ell>k$}.   
\end{cases}
\end{equation}
Additionally,
\begin{equation}
\label{eq:codesize}
\Pr\parenv*{\bX\in \bigwedge_{i\in [n-2k-\ell+2]} \overline{A_i}} = \Pr\parenv*{\bX\in\cC_{k,\ell}^{(0)}}=\frac{\abs{\cC_{k,\ell}^{(0)}}}{q^n}.
\end{equation}
   	
For the first case, assume $\ell\leq k$. Define
\[
c \eqdef \frac{1 - \sqrt{1 - \frac{\ell!(16k+8\ell-16)}{q^{k+\ell-1}}}}{8k+4\ell-8},
\]
and for all $i\in[n-2k-\ell+2]$ set $c_i=c$. We then have for all $i$,
\[
c_i \prod_{(i,j)\in E}(1-c_j)  \overset{(a)}{\geq} c(1-c)^{4k+2\ell-4} \overset{(b)}{\geq} c(1-(4k+2\ell-4)c) = \frac{\ell!}{q^{k+\ell-1}}\overset{(c)}{\geq} \Pr(A_i),
\]
where $(a)$ follows from the fact that each vertex is connected to at most $4k+2\ell-4$ other vertices, $(b)$ follows from Bernoulli's inequality $(1-x)^r \geq 1-rx$ for $r\geq 1$ and $x\in (0,1)$, and $(c)$ follows from~\eqref{eq:prup}. Note that $q^{k+\ell-1} \geq  \ell! (16k +8\ell -16)$ for any $q\geq \ceil{(\ell!(16k+8\ell-16))^{\frac{1}{k+\ell-1}}}$, which makes sure that $c_i\in(0,1)$. 
   
By Lemma~\ref{lem:local} and~\eqref{eq:codesize}, we have 
\[\abs*{\cC_{k,\ell}^{(0)}} \geq q^n (1 - c)^{n-2k-\ell+2},\]
and therefore,
\begin{align*}
\lim_{n\to \infty} R(\cC_{k,\ell}^{(0)}) \geq \lim_{n\rightarrow \infty} \frac{n + (n-2k-\ell+2)\log_q \parenv{1-c}}{n} = 1 - \log_q \parenv*{\frac{1}{1-c}}.
\end{align*}

For the second case, when $k < \ell$, we choose 
\[c \eqdef \frac{1 - \sqrt{1 - \frac{k!(16k+8\ell-16)}{q^{2k-1}}}}{8k+4\ell-8},\]
and $q\geq \ceil{(k!(16k+8\ell-16))^{\frac{1}{2k-1}}}$. The remainder of the proof is the same.
\end{IEEEproof}

\begin{remark}
\label{rem:bound}
While requirement on the alphabet size $q$ in Theorem~\ref{thm:lowbound} is modest, we may wish to remove it to obtain a lower bound on the asymptotic rate for small alphabet sizes. We can do so, at the price of a lower rate guarantee, in the following fashion. We can construct sequences $\bx\in\cC_{k,\ell}^{(0)}$ by picking arbitrary symbols in all locations except those that are $0$ modulo $k$. In those positions, say $i$, we choose $x_i$ in a way that $R_{i,\ell}(\bx)\neq R_{i-k,\ell}(\bx)$. This removes at most one possible letter from the alphabet, per such position. We then get for all $q\geq 2$,
\[
\lim_{n\to\infty}R(C_{k,\ell}^{(0)}) \geq 1-\frac{1}{k}\log_q\parenv*{\frac{q}{q-1}}.
\]
\end{remark}

The results of this section are summarized in Table~\ref{tab:unbounded}. Table~\ref{tab:values} shows the bounds for common values associated with nanopore sequencing.

\begin{table*}
\caption{The asymptotic rate of the optimal code correcting any number
of tandem duplication of length $k$ in the nanopore $\ell$-read vector}
\label{tab:unbounded}
\begin{center}
\renewcommand{\arraystretch}{2}
\begin{tabular}{c|c|l|l}
\hline
\hline
$k$ & $\ell$ & Asymptotic Rate & Comment \\
\hline
\hline
$k\geq 2$ & $1$ & $= \log_q \lambda_{k-1,q}$ & \cite[(4)]{jain2017duplication} (see also~\eqref{eq:gamma})\\
\hline
$1$ & $\ell\geq 2$ & $=\log_q (q-1)$ & Theorem~\ref{th:k1} \\
\hline
\multirow{5}{*}{$k\geq 2$} & \multirow{5}{*}{$\ell\geq 2$} &
$= \log_q \lambda_{k+\ell-2,q}$ & Theorem~\ref{thm:upbound}, $\ell|k$ (see also~\eqref{eq:gamma} and Remark~\ref{rem:alpha})\\
\cline{3-4}
& & $\leq \log_q \lambda_{k+\ell-2,q}$ & Theorem~\ref{thm:upbound} (see also~\eqref{eq:gamma} and Remark~\ref{rem:alpha})\\
\cline{3-4}
& & $\begin{array}{l} \geq 1-\log_q\parenv*{\frac{1}{1-c}} \text{ where}\\ c = \frac{1 - \sqrt{1 - \frac{\ell!(16k+8\ell-16)}{q^{k+\ell-1}}}}{8k+4\ell-8} \end{array}$ & Theorem~\ref{thm:lowbound}, $\ell\leq k$, $q\geq \ceil*{(\ell!(16k+8\ell-16))^{\frac{1}{k+\ell-1}}}$  \\
\cline{3-4}
& & $\begin{array}{l} \geq 1-\log_q\parenv*{\frac{1}{1-c}} \text{ where}\\ c = \frac{1 - \sqrt{1 - \frac{k!(16k+8\ell-16)}{q^{2k-1}}}}{8k+4\ell-8} \end{array}$ & Theorem~\ref{thm:lowbound}, $\ell> k$, $q\geq \ceil*{(k!(16k+8\ell-16))^{\frac{1}{2k-1}}}$  \\
\cline{3-4}
& & $\geq 1-\frac{1}{k}\log_q\parenv*{\frac{q}{q-1}}$ & Remark~\ref{rem:bound}  \\
\hline
\hline
\end{tabular}
\end{center}
\end{table*}

\begin{table}
\caption{The asymptotic rate of the optimal code correcting any number
of tandem duplication of length $1\leq k\leq 9$ in the nanopore $\ell$-read vector, $\ell=5,9$, over an alphabet of size $q=4$ (numerical values are rounded to $6$ decimal digits}
\label{tab:values}
\begin{center}
\begin{tabular}{c|c|c|c}
\hline
\hline
$k$ & $\ell$ & Lower Bound & Upper Bound \\
\hline
\hline
$1$ & $5$ & \multicolumn{2}{c}{$=0.792481$} \\
$2$ & $5$ & $0.896241$ & $0.999868$ \\
$3$ & $5$ & $0.995182$ & $0.999967$ \\
$4$ & $5$ & $0.998906$ & $0.999992$ \\
$5$ & $5$ & \multicolumn{2}{c}{$=0.999998$} \\
$6$ & $5$ & $0.999917$ & $0.999999$ \\
$7$ & $5$ & $0.999979$ & $1.000000$ \\
$8$ & $5$ & $0.999995$ & $1.000000$ \\
$9$ & $5$ & $0.999999$ & $1.000000$ \\
\hline
\hline
$1$ & $9$ & \multicolumn{2}{c}{$=0.792481$} \\
$2$ & $9$ & $0.896241$ & $0.999999$ \\
$3$ & $9$ & $0.994779$ & $1.000000$ \\
$4$ & $9$ & $0.998891$ & $1.000000$ \\
$5$ & $9$ & $0.999664$ & $1.000000$ \\
$6$ & $9$ & $0.999875$ & $1.000000$ \\
$7$ & $9$ & $0.999946$ & $1.000000$ \\
$8$ & $9$ & $0.999973$ & $1.000000$ \\
$9$ & $9$ & \multicolumn{2}{c}{$=1.000000$} \\
\hline
\hline
\end{tabular}
\end{center}
\end{table}

\section{Correcting a Constant Number of Errors}
\label{sec:bounded}

In this section we turn to look at the case of a constant number of tandem-duplication error of length $k$. As discussed in the introduction, we shall be paralleling a construction of such a code, bringing it to the domain of $\ell$-read vectors. Unlike the previous section, we shall parallel the construction given in~\cite{kovavcevic2018asymptotically}. As in the previous section, the challenge is not the construction itself, but rather finding its redundancy, and whether it is optimal. 

The construction of~\cite{kovavcevic2018asymptotically}, which we parallel here, uses Sidon sets. Let $G$ be a finite Abelian group. A set $B=\set{b_1,\ldots,b_r}\subseteq G$ is said to be a \emph{Sidon set of order $t$} if the sums $\sum_{i=1}^r a_i b_i$ have different values for all $a_1,\ldots,a_r\in \mathbb{Z}$ with $a_i\geq 0$ and $\sum_{i=1}^r a_i \leq t$. By the result of~\cite{bose1960theorems}, we can find such an Abelian group of size $\abs{G}=O(r^t)$, where we assume $r\to\infty$ while $t$ is fixed. More precisely, by~\cite[Theorem 2]{bose1960theorems}, for $r$ a prime power, we can take $G=\Z_q$ with $q=(r^{t+1}-1)/(r-1)$.

Let us introduce a useful notation. given two vectors of integers $\ba\in\Z^m$ and $\ba'\in\Z^{m'
}$, $m,m'\geq 1$, not necessarily of the same length, we define their dot product as
\[
\ba\odot\ba'\eqdef \sum_{i=1}^{\min(m,m')} a_i a'_i.
\]
With this notation we can proceed with the construction.

\begin{construction}
\label{con:bounded}
Let $k$, $\ell$, $t$, $n$, and $q\geq 2$ be positive integers. Let $G$ be an Abelian group containing a Sidon set $B=\set{b_1,\dots,b_{n+\ell}}$ of order $t$. Define $\bb\eqdef(b_1,\dots,b_{n+\ell})$. For any $g\in G$ we construct the code
\[
\cC_{k,\ell}^{g}(n) \eqdef \set*{ \bx\in\Z_q^n ~:~ \bb\odot\sigma_k(\Delta_k(\cR_\ell(\bx))) = g},
\]
where we recall the definition of $\sigma_k(\cdot)$ from~\eqref{eq:sigma}.
\end{construction}

\begin{theorem}
\label{th:codebounded}
The code $\cC_{k,\ell}^g(n)$ from Construction~\ref{con:bounded} can correct any $t$ or fewer tandem duplications of length $k$ in the $\ell$-read vectors of its codewords.
\end{theorem}

\begin{IEEEproof}
We prove the claim by giving a decoding procedure. Assume $\bx\in\cC_{k,\ell}^{g}(n)$ was transmitted, and denote its $\ell$-read vector by $\bz=\cR_\ell(\bx)$. By construction, \[\bb\odot\sigma_k(\Delta_k(\bz))=g.\]
Assume, however, that $\bz'$ is the $\ell$-read vector obtained at the channel output, $\bz\der_k^{i} \bz'$, where $0\leq i\leq t$. We can compute
\[
\bb\odot\sigma_k(\Delta_k(\bz'))=g'.
\]
By Lemma~\ref{lem:mu}, $\sigma_k(\Delta_k(\bz'))-\sigma_k(\Delta_k(\bz))$ contains only non-negative entries, and 
\[\norm{\sigma_k(\Delta_k(\bz'))-\sigma_k(\Delta_k(\bz))}_1=i.\]
Thus, by the properties of the Sidon set, knowing the value of the element
\[
g'-g = \bb\odot(\sigma_k(\Delta_k(\bz'))-\sigma_k(\Delta_k(\bz)))
\]
gives us exactly the entries of $\sigma_k(\Delta_k(\bz'))-\sigma_k(\Delta_k(\bz))$. We can therefore recover $\sigma_k(\Delta_k(\bz))$, which together with $\mu_k(\Delta_k(\bz'))=\mu_k(\Delta_k(\bz))$ immediately allows us to recover $\bz$.
\end{IEEEproof}

\begin{corollary}
\label{cor:red}
Let $k$, $\ell$, $t$, $n$, and $q\geq 2$ be positive integers. Then there exists a code $\cC$ of length $n$ over $\Z_q$ that can correct any $t$ or fewer tandem duplications of length $k$ in the $\ell$-read vectors of its codewords, and has redundancy $t\log_q n + O(1)$.
\end{corollary}

\begin{IEEEproof}
By~\cite{bose1960theorems}, there exists a Sidon set of order $t$ with $n+\ell$ elements over an Abelian group $G$ of size $\abs{G}=O(n^t)$. Consider the codes $\cC_{k,\ell}^g$ from Construction~\ref{con:bounded}. By Theorem~\ref{th:codebounded} they all possess the desired error-correction capability.

We note that
\[
\bigcup_{g\in G}\cC_{k,\ell}^g(n)=\Z_q^n.
\]
By averaging, there exists $g\in G$ such that
\[
\abs*{\cC_{k,\ell}^g(n)}\geq \frac{\abs{\Z_q^n}}{\abs{G}}=\frac{q^n}{O(n^t)}.
\]
By definition this code has redundancy $t\log_q n + O(1)$.
\end{IEEEproof}

We now turn to lower bound the redundancy required for a code that is capable of correcting $t$ tandem duplications of length $k$ in the $\ell$-read vector. Unlike the upper bound obtained through Construction~\ref{con:bounded}, it appears to be difficult to get a lower bound. We do so for the case of $\ell|k$, and show that it matches the upper bound of Corollary~\ref{cor:red} up to an additive number of $O(1)$ symbols.

\begin{theorem}
\label{th:boundedlow}
Let $k$, $\ell$, $t$, $n$, and $q\geq 2$ be positive integers. Assume that $\ell|k$. Then any code $\cC$ of length $n$ over $\Z_q$ that can correct any $t$ or fewer tandem duplications of length $k$ in the $\ell$-read vectors of its codewords, has redundancy at least $t\log_q n + O(1)$.
\end{theorem}

\begin{IEEEproof}
Assume $\cC$ is a code as stated in the theorem. Then $\cR_\ell(\cC)\eqdef\set{\cR_\ell(\bx) ~:~ \bx\in\cC}$ is a set capable of correcting up to $t$ duplications of length $k$. By the previous discussion, this means that $\Delta_k(\cR_\ell(\cC))\eqdef\set{\Delta_k(\cR_\ell(\bx)) ~:~ \bx\in\cC}$ is a code capable of correcting up to $t$ insertions of $0^k$. By~\cite[Lemma 1]{kovavcevic2018asymptotically}, $\Delta_k(\cR_\ell(\cC))$ can correct up to $t$ deletions of $0^k$.

For us to use the deletion-correction capability stated above, we need to bound the number of sequences $\bx\in\Z_q^n$ such that $\Delta_k(\cR_\ell(\bx))$ contains sufficiently many maximal runs of zeros of length at least $k$ (which allow a deletion of $0^k$ to occur). At this point we start using the assumption that $\ell|k$, and we make the following observations. Assume $\bx\in\Z_q^n$ is a sequence such that $0^k$ occurs at position $i+\ell+k-1$ of $\Delta_k(\cR_\ell(\bx))$. By Lemma~\ref{lem:0k}, $\bx[i,i+2k+\ell-2]$ is $k$-periodic. This means that $\Delta_k(\bx)$ (notice that we take the derivative of $\bx$ here, and not of the $\ell$-read vector of $\bx$) contains $0^{k+\ell-1}$ at position $i+k$. Since the other direction is trivial, we get
\begin{gather}
\Delta_k(\cR_\ell(\bx))[i+k+\ell-1,i+2k+\ell-2]=0^k \nonumber \\
\Updownarrow\nonumber\\
\bx[i,i+2k+\ell-2] \text{ is $k$-periodic}\nonumber\\
\Updownarrow\nonumber\\
\Delta_k(\bx)[i+k,i+2k+\ell-2]=0^{k+\ell-1}. \label{eq:cong}
\end{gather}
We also note that since all relevant transformations are invertible,
\[
\abs*{\Z_q^n}=\abs*{\Delta_k(\Z_q^n)}=\abs*{\Delta_k(\cR_\ell(\Z_q^n))}=q^n.
\]
Another crucial fact is that, when $\ell|k$, if we delete the $0^k$ factor from $\Delta_k(\cR_\ell(\bx))$, what remains is indeed a $k$-step derivative of an $\ell$-read vector of some sequence, specifically,
\begin{equation}
\label{eq:del}
\Delta_k(\cR_\ell(M_{[i,i+k-1]}(\bx)))=M_{[i+k+\ell-1,i+2k+\ell-2]}(\Delta_k(\cR_\ell(\bx)).
\end{equation}

We shall need the following notation. Assume $\Sigma$ is some finite alphabet that contains the symbol $0$. If $\bx\in\Sigma^n$ is some sequence, we can uniquely factor in the following way:
\[
\bx=0^{m_0} a_1 0^{m_1} a_2 0^{m_2} \dots 0^{m_{s-1}} a_s 0^{m_s},
\]
for some non-negative integers $m_i$, and where $a_i\in\Sigma\setminus\set{0}$. We say $\bx$ contains $m$ maximal runs of zeros of length $\geq r$ if there are exactly $m$ indices $i$ for which $m_i\geq r$. The set of all such sequences is denoted by $S_m^{\geq r}(\Sigma^n)$. Back to our problem, we shall be interested in the size of $S_m^{\geq k+\ell-1}(\Z_q^n)$.

By~\cite[Lemma 3]{kovavcevic2018asymptotically}\footnote{We mention that~\cite[Lemma 3]{kovavcevic2018asymptotically} contains an extra weight restriction which is of no consequence here.}, there exists a sub-linear function $f(n)=o(n)$ such that
\[
\sum_{\abs{m-\mu n}>f(n)} \abs*{S_m^{\geq k+\ell-1}(\Z_q^n)} \leq \frac{q^n}{n^{\log_2 n}},
\]
where $\mu\eqdef\frac{q+1}{q^{k+\ell}}$. Loosely speaking, most sequences contain around $\mu n$ maximal runs of zeros of length $\geq k+\ell-1$. By~\eqref{eq:cong}, we have
\begin{equation}
\label{eq:part1}
\sum_{\abs{m-\mu n}>f(n)} \abs*{S_m^{\geq k}(\Delta_k(\cR_\ell(\Z_q^n))} \leq \frac{q^n}{n^{\log_2 n}},
\end{equation}

Let $\cC'\subseteq \bigcup_{\abs{m-\mu n}\leq f(n)} S_m^{\geq k}(\Delta_k(\cR_\ell(\Z_q^n))$ be a code capable of correcting up to $t$ deletions of $0^k$. Since each $\bx\in\cC'$ contains at least $\mu n-f(n)$ maximal runs of zeros of length $\geq k$, after exactly $t$ deletions of $0^k$, the number of resulting sequences is at least $\binom{\mu n - f(n)}{t}$. Additionally, by~\eqref{eq:del}, after these $t$ deletions the sequence resides in $\Delta_k(\cR_\ell(\Z_q^{n-tk}))$. Since after the deletions, all sequences must be distinct,
\[
\abs*{\cC'}\binom{\mu n-f(n)}{t} \leq \abs*{\Delta_k(\cR_\ell(\Z_q^{n-tk}))}.
\]
Rearranging, and using $\binom{a}{b}\geq a^b/b^b$,
\begin{equation}
\label{eq:part2}
\abs*{\cC'} \leq \frac{q^{n-tk} t^t}{(\mu n- f(n))^t}.
\end{equation}

Finally, assume $\cC\subseteq\Z_q^n$ is a code satisfying the conditions of the theorem. We have $\abs{\cC}=\abs{\Delta_k(\cR_\ell(\cC))}$. The latter we partition into two parts:
\begin{align*}
\cC'_1&\eqdef \Delta_k(\cR_\ell(\cC))\cap \bigcup_{\abs{m-\mu n}\leq f(n)} S_m^{\geq k}(\Delta_k(\cR_\ell(\Z_q^n)), \\
\cC'_2&\eqdef \Delta_k(\cR_\ell(\cC))\cap \bigcup_{\abs{m-\mu n}> f(n)} S_m^{\geq k}(\Delta_k(\cR_\ell(\Z_q^n)).
\end{align*}
By~\eqref{eq:part1} and~\eqref{eq:part2} we have
\begin{align*}
\abs*{\cC'_1} &\leq  \frac{q^{n-tk} t^t}{(\mu n- f(n))^t},\\
\abs*{\cC'_2} &\leq \frac{q^n}{n^{\log_2 n}}.
\end{align*}
Thus,
\[
\abs*{\cC}=\abs*{\cC'_1}+\abs*{\cC'_2}\leq \frac{q^{n-tk} t^t}{(\mu n- f(n))^t} + \frac{q^n}{n^{\log_2 n}}=O\parenv*{\frac{q^n}{n^t}},
\]
and the redundancy of $\cC$ is at least $t\log_q n + O(1)$.
\end{IEEEproof}

\begin{remark}
We note that if we remove the restriction of $\ell|k$, the proof of Theorem~\ref{th:boundedlow} may break. For example, we show a counter-example to~\eqref{eq:del}. Assume $q=2$, $\ell=2$, $k=3$, and $n=7$. Consider the sequence $\bx=(0,1,0,1,0,1,0)$. Then,
\begin{align*}
\cR_2(\bx)&=(\sz_0,\sz_0+\sz_1,\sz_0+\sz_1,\sz_0+\sz_1,\sz_0+\sz_1,\sz_0+\sz_1,\sz_0+\sz_1,\sz_0),\\
\Delta_3(\cR_2(\bx))&=(\sz_0,\sz_0+\sz_1,\sz_0+\sz_1,\sz_1,0,0,0,-\sz_1).
\end{align*}
Let us now delete the $0^3$ factor in $\Delta_3(\cR_2(\bx))$ to obtain
\[
\bz'=(\sz_0,\sz_0+\sz_1,\sz_0+\sz_1,\sz_1,-\sz_1).
\]
If we compute $\Delta_3^{-1}(\bz')$ we get
\[
\Delta_3^{-1}(\bz')=(\sz_0,\sz_0+\sz_1,\sz_0+\sz_1,\sz_0+\sz_1,\sz_0).
\]
However, $\Delta_3^{-1}(\bz')$ is not the $2$-read vector of any binary sequence. Thus, $\bz'\not\in\Delta_3(\cR_2(\Z_2^4))$.
\end{remark}

\section{Conclusion}
\label{sec:conc}

In this paper we studied the asymptotic rate of the optimal code capable of correcting tandem duplications of length $k$ in the $\ell$-read vector of the nanopore channel. The number of correctable tandem duplications was either unbounded (in Section~\ref{sec:unbounded}), or bounded by a constant (in Section~\ref{sec:bounded}).

We observe that in the case of unbounded tandem duplication errors, the redundancy is linear in the length of the code $n$. This channel generalizes the channel described in~\cite{jain2017duplication}. Both the lower and upper bounds tend to $1$ exponentially fast as $\ell,k\to\infty$. As Table~\ref{tab:values} shows, even for practical values, the rate is nearly $1$.

When the number of tandem duplications is bounded by $t$ errors and $\ell|k$, our bounds match asymptotically, giving a redundancy of $t\log_q n+O(1)$. It is interesting to note that this expression does not depend on $\ell$ and $k$, which agrees with the special case ($\ell=1$) studied in~\cite{kovavcevic2018asymptotically}. However, when $\ell\nmid k$, we only have $t\log_q n+O(1)$ as an upper bound on the redundancy.

Some open questions remain. In particular, we are interested in finding explicit constructions for asymptotically optimal codes. The special case of $\ell=1$ studied for unbounded number of errors in~\cite{jain2017duplication} does provide an explicit construction. In our case, there seems to be an interesting connection between the parameters $k$ and $\ell$, making it difficult to give an explicit construction similar to~\cite{jain2017duplication}. Thus, while the construction we describe is optimal, more work is required to flesh out an explicit one. When we restrict ourselves to a bounded number of errors, already~\cite{kovavcevic2018asymptotically} has an optimal construction, but that one is not explicit. Finding such constructions is left for future work.

\bibliographystyle{IEEEtranS}
\bibliography{reference}

\end{document}